\documentstyle[12pt, aasms4, epsfig]{article}
\catcode`@=11
\def\eqalign#1{\null\,\vcenter{\openup\jot\m@th
  \ialign{\strut\hfil$\displaystyle{##}$&$\displaystyle{{}##}$\hfil
      \crcr#1\crcr}}\,}
\def\eqalignleft#1{\null\,\vcenter{\openup\jot\m@th
  \ialign{\strut$\displaystyle{##}$\hfil&$\displaystyle{{}##}$\hfil
      \crcr#1\crcr}}\,}
\catcode`@=12

\def\lax    {\ifmmode{_<\atop^{\sim}}\else{${_<\atop^{\sim}}$}\fi}
\def\gax    {\ifmmode{_>\atop^{\sim}}\else{${_>\atop^{\sim}}$}\fi}
\def\kms    {\ifmmode{{\rm ~km~s}^{-1}}\else{~km~s$^{-1}$}\fi}

\def\bk{\lower 6pt\hbox{${\buildrel k\over \sim}$}}
\def\bv{\lower 6pt\hbox{${\buildrel v\over \sim}$}}

\begin{document}
\textwidth 6.5truein
\textheight 9.25truein
\topmargin -1cm
\title{\bf 
The Large Scale Structures in the Solar System:\\
I. Cometary Belts With Resonant Features\\
Near the Orbits of Four Giant Planets
}
\author{\bf Leonid M. Ozernoy\altaffilmark{1}}  
\affil{5C3, Computational Sciences  Institute and Department of Physics 
\& Astronomy,\\ George Mason U., Fairfax, VA 22030-4444; also Laboratory for 
Astronomy and Solar\\ Physics, NASA/Goddard Space Flight Center, Greenbelt, 
MD 20771} 
\altaffiltext{1}{e-mail: ozernoy@science.gmu.edu; ozernoy@stars.gsfc.nasa.gov}
\author{\bf Nikolai N. Gor'kavyi\altaffilmark{2}}
\affil{
Laboratory for Astronomy and Solar Physics, NASA/Goddard Space Flight 
Center\\ Greenbelt, MD 20771; also 
Simeiz Department, Crimean Astrophysical Observatory, Simeiz 334242, Ukraine
}
\altaffiltext{2}{NRC/NAS Senior Research Associate; e-mail: 
gorkavyi@stars.gsfc.nasa.gov}
\author{\bf Tatiana Taidakova\altaffilmark{3}}
\affil{
Simeiz Department, Crimean Astrophysical Observatory, Simeiz 334242, Ukraine
}
\altaffiltext{3}{e-mail: gorkavyi@stars.gsfc.nasa.gov}
\bigskip

\begin{abstract}
We employ an efficient numerical approach to simulate a stationary distribution
of test objects, which results from their gravitational scattering on the 
four giant planets, with accounting for effects of mean motion resonances. 
Using the observed distribution of the Kuiper belt objects, we reconstruct, 
in the space of orbital coordinates, the distribution function $n (a,e,i)$ 
for the population of minor bodies beyond Jupiter. We confirm that thousands 
of large yet cold comets and Centaurs might be located between the orbits of 
Jupiter and Neptune. Moreover, we find as an important result that they are 
concentrated into four circumsolar belts, with a highly non-uniform and well 
structured distribution of the objects. This huge yet unrevealed population, 
with only a few of its representatives presently known, is expected to have, 
like our simulations demonstrate, a rich resonant structure containing both 
density maxima and gaps. The resonant structure is formed due to 
gravitational perturbations, i.e. in a non-dissipative way. If plotted in the 
($a,e,i$)-space of orbital coordinates, the belts contain gaps (including 
those between resonant groups), quite similar to the Kirkwood gaps in the 
main asteroid belt. An appreciable fraction  of the test bodies reveals, 
for some time, an accumulation near (rather than in) the resonances, both 
interior and exterior, with the giant planets. 

An accompanying paper considers the population simulated in this work as the 
major source of dust in the outer Solar system. The simple but fast and 
efficient numerical approach employed in this work would allow the reader 
for applying it to many other problems of his/her interest. 

\end{abstract}
\newpage
\section*{1. Introduction}
Not much is known about the size of the population of minor bodies of the  
Solar System between the 
orbits of Jupiter and Neptune, nor how these bodies are distributed there. 
The known representatives of that population, the Centaurs, are the objects 
with $6\lax a\lax 26$ AU intermediate between comets and asteroids, having 
presumably short lifetimes compared to the Solar system's age (Asher \&
Steel 1993). These objects, as well as the other known populations, such 
as the Jupiter-family comets, could have a common origin in the Kuiper belt  
(Luu \& Jewitt 1996). 
For Jupiter-family comets, an alternative evolutionary path from the Oort
cloud has been found unlikely (see e.g. review by Kres\'ak 1994 and refs.
therein). If the Kuiper belt objects (called `kuiperoids' hereinafter)
are indeed responsible for progressive replenishment of the observable  
populations, gravitational 
scattering on all four giant planets would be responsible for the transport
of these objects from the trans-Neptunian region all the way inward, down to 
Jupiter (Levison \& Duncan, 1997). The present paper aims at a related, 
but rather different subject, {\it viz.} computing
the distributions of minor bodies between  Neptune and Jupiter, including
analysis of those distributions in the space of orbital coordinates, $a,e,i$.
An accompanying paper deals with the distribution of dust expected to be
produced by these bodies.

\section*{2. Processes Accounted for: Gravitational Scattering and 
Influence of Resonances}

Gravitational scattering of test bodies on the planets 
(Carusi, Valsecchi, \& Greenberg, 1990) and influence of resonances 
(Jackson \& Zook 1989, Roques et al. 1994) 
have been studied previously by a number of investigators. Here, we undertake 
a combined numerical study of these effects to demonstrate that,
when put together, they mutually enhance each other by creating
qualitatively new phenomena. 
While using a different numerical approach and employing different tools
to analyze the computational results, we also make
emphasis on a different subject -- the stationary distributions of
comets near each of the giant planets and its structure in the space
of orbital coordinates. [A systematic use of distributions in these 
coordinates is a part of the kinetic approach employed by us in studying 
the structure, evolution, and origin of the interplanetary dust and its
sources (Gor'kavyi et al. 1997a,b, 1998)]. Furthermore, as opposed to the
usual consideration of a {\it dissipative} mechanism of captures
into mean motion  resonances (e.g., due to the tidal dissipation or
Poynting-Robertson drag), we deal here with an entirely different type
of {\it non-dissipative} captures, when the Tisserand constant
is conserved. Indeed, gravitational scattering as the 
only accompanying process in our consideration here is an elastic
process.
A dissipationless capture of a minor body into a mean motion resonance 
with the planet becomes possible at a large eccentricity of a minor body, 
when an exchange of angular momentum
between the planet and the minor  body becomes especially 
effective due to their very close approach.

In our numerical study, we adopt the approximations of a restricted
3-body problem (the Sun, the planet on a circular orbit, and a massless 
minor body).
It is convenient to define the planet's zone of gravitational influence  
in the $(a,e)$-plane of orbital coordinates:
\begin{eqnarray}
a(1-e)&\leq & a_p ~~~{\rm if}~ a>a_p,\\
a(1+e)&\geq & a_p ~~~{\rm if}~ a<a_p,
\end{eqnarray}
where $a$ is the semi-major axis of a test body,
$a_p$ is the semi-major axis of the planet, and
$e$ is eccentricity of the test body. This zone looks like a triangle
and we call it hereinafter `the triangle zone'.

Our study examines the distribution of minor bodies within each 
planet's zone of influence. We also evaluate qualitatively 
how the bodies are scattered from one zone to another.

Obviously, a restricted 3-body problem is a highly simplified approach
to examine the distribution of minor bodies in the gravitational field
produced by the four giant planets. However, as the reader will see
below, even this approximation reveals a rich and sophisticated structure
in the distribution of test bodies. Our follow-up research 
will study how the revealed distribution is modified when 
the influence of three other planets, with accounting for non-circular 
orbits as well, is taken into consideration.
\section*{3. Computational Method: Simulation of a Quasi-stationary 
Distribution of Test Bodies in the Planet's Zone of Influence}

A quasi-stationary distribution of minor bodies has been established
in the outer parts of the Solar system due to multiple 
gravitational scatterings on the four giant planets.
Each giant planet maintains this quasi-stationarity in its 
zone of gravitational influence by ejecting the minor bodies in the 
amounts comparable to what flows into this zone from outside.
As a convenient approach to simulate such a quasi-stationary
distribution of massless minor bodies around a giant planet, we applied 
the following computational procedure: a record of coordinates and velocities
of a test body was taken after certain number of revolutions (usually each 
10 revolutions) of the planet
around the Sun and these data were then used to characterize the positions 
of {\it many} test bodies over the entire time span, beginning at
an initial instant, and ending at the instant of ejection of the test body 
from the planet's zone of influence.

We computed 44 stationary distributions of test bodies totalling $0.5\times
10^6$ $(a,e,i)$-orbital elements. Details of 
computational runs are given in Table 1.
\begin{center}
{\%\% PUT TABLE 1 HERE\%\%}
\end{center}

Meanwhile a small fraction of minor bodies is 
scattered from the zone of influence of each outer 
giant planet into the next innermost 
giant neighbor's zone. As a result, there is a 
flow of these objects from the trans-Neptune region
inward the Solar system. 

\section*{4. The Results: Four Cometary-Asteroidal 
Belts and Their Resonant Structure}

Our results of orbit integrations are shown, in the orbital
coordinates $a,e$ and $a,i$, for Neptune in Fig.~1a,b; for Uranus in 
Fig.~3a,b;
for Saturn in Fig.~5a,b; and for Jupiter in Fig.~7a,b.


\subsection*{4.1. Four Cometary-Asteroidal Belts} 

The minor bodies (kuiperoids and Centaurs) from the trans-Neptune
regions can be captured
into  mean motion resonances with Neptune  (Morbidelli 1997, Levison \& 
Dunkan 1997, Malhotra 1998). 
A minor body being 
in a resonance with Neptune increases its eccentricity, approaches closely 
the planet
and undergoes a strong gravitational scattering on it. The scattering
results in filling a quasi-triangle zone with numerous resonances 
and gaps  (see Fig.~1a). 

The minor bodies populating Neptune's zone of gravitational influence
should form a broad belt along Neptune's orbit -- the `Neptune belt'.
Furthermore,
a fraction of minor bodies from Neptune's quasi-triangle zone is 
intercepted by Uranus and fills in the quasi-triangle zone of the latter 
in the same fashion as it happens for Neptune's zone of influence.
A similar process of `leakage' of a part of minor bodies occurs from the 
belt of each outer  giant planet into the next innermost 
giant neighbor's sphere of influence. Some of them (e.g. Uranus) are
able to eject, by gravitational scattering, a minor body with such a 
high velocity in a particular direction that the body can reach Jupiter.

As a net result, there is a flow of these objects from the trans-Neptune 
region downward to Jupiter.  The minor bodies form a rather wide belt around 
each giant planet's orbit. Figs. 1a, 3a, 5a, and 7a
indicate that the bulk of test bodies is located within each planet's
triangle zone of influence, although a substantial number of bodies
form a `beard-like' distribution with lower eccentricities.

\subsection*{4.2. Resonant Structure of the Four Belts}

Two types of structures in the zone of gravitational influence of each 
giant planet are clearly seen in Figs. 1 to 8: (i) resonant gaps
and (ii) resonant groups of test bodies. There is a substantial difference 
between them: gaps represent a robust feature, whereas probability
to be captured into a resonant group is  rather sensitive to the number
of computational runs used.

We would like to emphasize an interesting phenomenon of a {\it
near-resonance accumulation}: the test bodies
tend to avoid location directly within the resonances, both
interior and exterior ones. Instead, they prefer to be positioned slightly
aside, mostly at both sides of the resonant gaps. Those gaps seem to be
similar to the well-known Kirkwood gaps in the main 
asteroid belt. 

The locations of gaps are clearly seen from
 the number of test bodies as a function
of semi-major axis, which is shown, for each of the giant planets, 
in Figs.~2, 4, 6, and 8. Occurence of locations is a measure of a probability
for a test body, $p(a)$, to be located between $a$ and $a+da$ (with $a$
measured in $a_{planet}$ , where $da=0.05$ and the total probability 
to be in the planet's triangle zone is $\int p(a)~da=1$.
 
There are two possible mechanisms for gap formation. The first one is
associated directly with gravitational scattering of a test body 
on the planet. The scattering can be stronger when the
minor body's period of revolutions is commensurable with the planet's period,
and this is not accompanied by the resonant capture thereby producing a gap.
The second mechanism is associated with the resonance influence: by getting 
an excessive angular momentum from the planet, the minor body leaves the 
resonance.

When the minor body leaves the resonance, the balance of the angular
momentum forces the minor body to librate, thereby
 spending much of the time at the edge of the resonant region. This results
in a {\it near-resonant accumulation}. We observe numerous examples of
such accumulations (see e.g. Figs.~3 and 5). Without going into a 
detailed classification, we will not discriminate here between usual
resonance and near-resonant accumulation.

Test bodies captured into resonances in the triangle  zone, undergo 
a characteristic pattern of evolution in $e$ and $i$ within 
the resonances:  
As long as the eccentricity of the minor body 
captured into a particular resonance increases (decreases), its inclination 
(shown in Figs.~1b, 3b, 5b, and 7b) decreases (increases) so that the Tisserand 
constant, $T$, is kept invariable. In our computations,
$T$ is only insignificantly (at the typical level of a fraction of 1\%) 
fluctuates. 


A detailed consideration of the dynamics leading to the capture 
into, as well as strength of, resonances when gravitational scattering
is the only accompanying process is considered elsewhere (Gor'kavyi \&
Ozernoy 1999).

\subsection*{4.3. The Expected General Picture of the Belts} 

For illustrative purposes, in Fig.~9 we show 
in $(a,e)$-coordinates the positions of all four cometary
belts we expect to exist near the orbits of the four 
giant planets. The known minor bodies in the main asteroid belts
and the known Centaurs are also shown Fig.~9. Although this picture 
neglects eccentricities of the giant planets 
as well as gravitational perturbations from three other giant planets 
in each planetary zone of influence 
(thereby secular resonances are out of consideration 
here), the resonant patterns in each zone are
expected to be preserved. It is remarkable that, as is seen in Fig.~9, 
at least 4 of 6 known Centaurs are located close to the resonances.

Finally, Figs. 10 and 11 illustrate, in the above approximations, 
the large-scale structure of the Solar system formed by the 
cometary-asteroidal 
belts envisaged in this paper. The maxima in minor
body concentrations clearly delineate the positions of these belts,
although the actual density contrasts should be determined in further,
more detailed computations.

\section*{5. Conclusions}

1. As our simulations indicate, the distribution of cometary 
and asteroidal bodies between the orbits of Jupiter and Neptune
is expected to be structured into four belts. These belts are 
well separated in the $(a,e$)-plane of the orbital coordinates. Moreover, 
they are expected to be distinguishable in space  as well. 

2. The cometary-asteroidal belts near all four giant planets contain
a rich resonant structure. Each belt plotted in the orbital coordinate 
$(a,e,i)$-space, is characterized by sharp edges and an internal structure
that includes density maxima and gaps around resonances. The latter 
are similar to the Kirkwood gaps in the main asteroid belt.

3. The minor bodies near the boundary of (or inside) the planet's triangle 
zone described by Eqs.(1)-(2) can be captured into non-dissipative resonances.
More precisely, the bodies tend to be located not directly in the resonances
but, instead, they reveal a near-resonance accumulation on the both sides
of the resonance.
Dynamics of such bodies can be explained by a balance between resonant 
interactions with the planet and close approaches to it.

4. We expect that an appreciable part of newly discovered Centaurs 
between Jupiter and Neptune be in resonances 
with the appropriate planet and located
outside the `triangle zones' of the giant planets.

\vspace{0.15in}
The resonant structure of the cometary-asteroidal belts near the outer
planets simulated in this work has some general features which could
be revealed both in the distribution of AAA-asteroids near Earth and near
exo-planets.

The envisaged cometary-asteroidal belts, which represent the largest 
structures in 
the Solar system, are expected to serve as the major sources of dust in the
outer parts of our planetary system. The dust distribution produced by
these sources is the subject of our accompanying paper.

\vspace{0.1in}
{\it Acknowledgements.} This work has been supported by NASA Grant NAG5-7065 
to George Mason University. N.G. acknowledges the NRC-NAS associateship.
T.T. is thankful to the American Astronomical Society for a Small Research
Grant from the Gaposchkin's Research Fund.

\newpage
\centerline{TABLE 1}
\medskip
\hrule
\medskip
\centerline{Details of Computational Runs}
\medskip
\hrule
\smallskip
\hrule
\medskip
\centerline{\vbox{
\halign{ \hfil # \hfil  & \hfil # \hfil  & \hfil # \hfil
& \hfil # \hfil  & \hfil # \hfil 
\cr
  & Neptune  & Uranus & Saturn & Jupiter  \cr
\noalign{\vskip 10pt}
 Number of runs  & 3 & 5 & 16 & 20 \cr
\noalign{\vskip 10pt}
 Number of planet's       & from $0.500\times 10^6$  & from 
 $0.76\times 10^6$ & from $1.69\times 10^4$  & from $1.50\times 10^3$  \cr
revolutions during one run$(^1)$& to $0.985\times 10^6$    & to 
$3.0\times 10^6$ & to $3.92\times 10^5$ & to $0.59\times 10^5$ \cr  
\noalign{\vskip 10pt}
 Number of computed positions    & $198,500$ $(^2)$    & $103,495$ 
 $(^3)$ & $144,226$ $(^{2})$ & 36,421 $(^2)$ \cr
\noalign{\vskip 10pt}
 Number of positions   &  &  &  &  \cr
plotted in Figs. 1 to 4 & 29,885 $(^4)$ & 25,024 $(^5)$ & 28,848 $(^6)$ 
& 23,602 $(^7)$ \cr
\noalign{\vskip 8pt}
}}}
\hrule
\bigskip

$(^1)$ until the test body was ejected (otherwise the run was stopped).

$(^2)$ taken with time step $=$ 10 revolutions of the planet.

$(^3)$ taken with time step $=$ 100 revolutions of the planet.

$(^4)$ taken with time step $=$ 60 revolutions of the planet, with removing
the positions at $a>10~a_p$.

$(^5)$ taken with time step $=$ 400 revolutions of the planet, with removing
the positions at $a>10~a_p$.

$(^6)$ taken with time step $=$ 50 revolutions of the planet, with removing
the positions at $a>10~a_p$.

$(^7)$ taken with time step $=$ 10 revolutions of the planet, with removing
the positions at $a>10~a_p$.

\newpage
\centerline{\bf Figure Captions}
\vspace{0.1in}
{\bf Figure 1}.

{\bf a}.
Eccentricity vs. the semi-major axis of minor bodies in and near Neptune's 
zone of influence
from a simulation of 29,885 test body positions. 
The sides of the `triangle zone' given by Eqs. (1)-(2) are shown by dashed  
lines, and its base (the upper boundary of eccentricities of test bodies)
is determined by the Tisserand constant. Test bodies on   resonant and chaotic
orbits below the triangle zone's sides  delineate a `beard'. Both the
triangle zone and its `beard' contain 
numerous gaps and resonances, including those of a high order (e.g. 7:1, 
6:1, etc.)

{\bf b}.
Inclination vs. the semi-major axis of minor bodies in and near 
Neptune's zone from the same simulation.
Note an interesting two-side accumulation of test bodies around 4:3 resonance
as well as one-side accumulation of test bodies around 1:1 resonance.

\vspace{0.2in}
{\bf  Figure 2}. Occurence of various resonances and gaps in and near 
Neptune's zone from the same simulation as Fig.~1.

\vspace{0.2in}
{\bf Figure 3}.

{\bf a}.
Eccentricity vs. the semi-major axis of minor bodies in and near 
Uranus zone of influence from a simulation of 25,024 test body positions.

{\bf b}.
Inclination vs. the semi-major axis of minor bodies in and near Uranus zone
from the same simulation.
Note several `muffs' (i.e. symmetric accumulations localized within a short 
range of inclinations) near the resonances 2:1, 3:1, 6:1, and 7:1.

\vspace{0.2in}
{\bf Figure 4}.
 Occurence of various resonances and gaps in and near 
Uranus zone from the same simulation as Fig.~3.

\newpage
\vspace{0.2in}
{\bf Figure 5}.

{\bf a}.
Eccentricity vs. the semi-major axis of minor bodies in and near Saturn's
zone of influence 
from a simulation of 28,848 test body positions.
Numerous gaps and are clearly seen in the resonances 7:1, 13:2, 6:1, etc. 

{\bf b}.
Inclination vs. the semi-major axis of minor bodies in and near Saturn's zone
from the same simulation.
Note several `muffs'  similar to those seen in Fig.~2b.
They are formed 
at $a(1-e)\approx 1$, i.e. near the edge of the triangle zone.

\vspace{0.2in}
{\bf Figure 6}.
 Occurence of various resonances and gaps in and near 
Saturn's zone from the same simulation as Fig.~5.

\vspace{0.2in}
{\bf Figure 7}.
{\bf a}.
Eccentricity vs. the semi-major axis of minor bodies in and near Jupiter's 
zone of influence
from a simulation of 23,602 test body positions. One can see that the Jovian 
gaps have wider and less sharp edges compared to those of less massive giant
planets.

{\bf b}.
Inclination vs. the semi-major axis of minor bodies in and near Jupiter's zone
from the same simulation.
Note numerous `muffs' similar to those seen in Figs.~1b, 3b, and 5b.

\vspace{0.2in}
{\bf Figure 8}.
Occurence of various resonances and gaps in and near Jupiter's zone from the 
same simulation as Fig.~5. The structure of the resonant region 
between resonances 2:1 and
 4:1 looks very similar to the Kirkwood gap in the main asteroid belt
 between 1:4 and 1:2 resonances.

\vspace{0.2in}
{\bf Figure 9}. 
The large-scale structure of the outer part of the Solar system shown
in the orbital $(a,e)$-coordinates. The simulated cometary-asteroidal belts
are shown along with the currently known objects.
Crosses stand for asteroids of the main belt (100 objects), 
triangles stand for Jupiter-family comets (112 objects), 
squares stand for Centaurs (6 objects), and 
diamonds stand for kuiperoids (50 objects). The simulated population 
fills the densest parts of the triangle zones of all the giant planets.
The known Centaurs and those kuiperoids whose orbits cross the Neptune's
orbit are in this densest part as well. One can see that bulk of the known 
Jupiter-family comets are located far from the simulated, yet unrevealed,
cometary population in the densest part of Jupiter's triangle zone. We 
explain this by observational selection: the major part of the known 
Jupiter-family comets has perihelia $p<2$ AU (above the dashed line).  
The sides of the `triangle zone' given by Eqs. (1)-(2) are shown by heavy  
lines.

\vspace{0.2in}
{\bf Figure 10}. 
The large-scale structure of the outer part of the Solar system 
shown face-on.
For convenience, positions of Jupiter, Saturn, Uranus, and Neptune
are indicated. One can see (especially clearly for Neptune and Jupiter) 
the simulated cometary-asteroidal belts associated with the giant planet 
orbits.

\vspace{0.2in}
{\bf Figure 11}. 
The large-scale structure of the outer part of the Solar system 
shown   edge-on.
For convenience, positions of Jupiter, Saturn, Uranus, and Neptune
are indicated. One can see (especially clearly for Neptune and Jupiter) 
the simulated cometary-asteroidal belts associated with the giant planet 
orbits.

\newpage

\centerline{\bf References}
\def\ref#1  {\noindent \hangindent=24.0pt \hangafter=1 {#1} \par}
\smallskip
\ref{Asher, D.J. \& Steel, D.I. 1993, Mon. Not. Roy. Astr. Soc. 263-179}
\ref{Carusi, A., Valsecchi, G.B. \& Greenberg, R., 1990,
     Celest. Mech. 49, 111}
\ref{Gor'kavyi, N.N., Ozernoy, L.M. \& Mather, J.C. 1997a, ApJ 474, 496}
\ref{Gor'kavyi, N.N., Ozernoy, L.M., Mather, J.C. \& Taidakova, T. 
1997b, ApJ 488, 268}
\ref{ Gor'kavyi, N.N., Ozernoy, L.M., Mather, J.C. \& Taidakova, T.
   1998, Earth, Planets and Space, Vol. 50, 539}
\ref{Gor'kavyi, N.N., Ozernoy, L.M., 1999, ApJ  (to be submitted)}
\ref{Jackson, A.A. \& Zook, H.A. 1989, Nature 337, 629}
\ref{Kres\'ak, $\check{\rm L}$. 1994, in {\it Asteroids, Comets, Meteors 
1993} (A. Milani et al., eds.),  p. 77}
\ref{Levison, H.F. \& Duncan M.J. 1997, Icarus 127, 13}
\ref{Luu, J.X. \&  Jewitt, D.C. 1996, AJ 112, 2310}
\ref{Malhotra, R. 1998, in {\it Solar System Formation and Evolution"}
(D. Lazzaro et al. eds). ASP Conf. Ser. 149, 37} 
\ref{Marsden, B.G. 1998, MPEC 1998-v14: Distant Minor Planets}
\ref{Morbidelli, A. 1997, Icarus 127, 1}
     
\end{document}